\documentclass[runningheads]{llncs}

\usepackage{amsmath}
\usepackage{amsfonts}
\usepackage{graphicx}
\usepackage{float} 
\usepackage{tikz}
\usepackage{ctable}
\usepackage{comment}
\usepackage{makecell}
\usepackage{caption}
\usepackage{tabularx}
\usepackage{hyperref}
\usepackage{textgreek}
\usepackage{mathtools}
\usepackage{textcomp}
\usepackage{algorithm}
\usepackage{algorithmic}
\usepackage{makecell}
\usepackage{wasysym}
\usepackage{enumitem}
\usepackage{paralist}
\usepackage{MnSymbol}
\usepackage{tabulary}
\usepackage{threeparttable}
\usepackage{booktabs}
\usepackage{todonotes}
\usepackage{subcaption}
\usepackage[misc]{ifsym}
\hypersetup{
    colorlinks=true,
    linkcolor=red,
    filecolor=magenta,      
    urlcolor=red,
}

\newcommand{\vect}[1]{\boldsymbol{#1}}

\begin{document}

\title{Brain Tissue Segmentation Across the Human Lifespan via Supervised Contrastive Learning}

\titlerunning{Lifespan Brain Tissue Segmentation}

%
\authorrunning{Anonymous}


\author{Xiaoyang Chen, Jinjian Wu, Wenjiao Lyu, Yicheng Zou, Kim-Han Thung, Siyuan Liu, Ye Wu, Sahar Ahmad, and Pew-Thian Yap\textsuperscript{(\Letter)}}

\authorrunning{Chen et al.}

\institute{Department of Radiology and Biomedical Research Imaging Center (BRIC), \\University of North Carolina, Chapel Hill, NC, USA\\ 
{\Letter}~\email{ptyap@med.unc.edu}
}

\maketitle

\begin{abstract}
Automatic segmentation of brain MR images into white matter (WM), gray matter (GM), and cerebrospinal fluid (CSF) is critical for tissue volumetric analysis and cortical surface reconstruction. Due to dramatic structural and appearance changes associated with developmental and aging processes, existing brain tissue segmentation methods are only viable for specific age groups. Consequently, methods developed for one age group may fail for another. In this paper, we make the first attempt to segment brain tissues across the entire human lifespan ($0-100$ years of age) using a unified deep learning model. To overcome the challenges related to structural variability underpinned by biological processes, intensity inhomogeneity, motion artifacts, scanner-induced differences, and acquisition protocols, we propose to use contrastive learning to improve the quality of feature representations in a latent space for effective lifespan tissue segmentation. We compared our approach with commonly used segmentation methods on a large-scale dataset of 2,464 MR images. Experimental results show that our model accurately segments brain tissues across the lifespan and outperforms existing methods.
\end{abstract}

\section{Introduction}


Segmentation of gray matter (GM), white matter (WM), and cerebrospinal fluid (CSF) from magnetic resonance (MR) images is a prerequisite for delineating brain anatomical structures and for quantifying changes in tissue volumes and cortical geometry in relation to development and aging. Manual brain tissue  segmentation carried out by experts can be time-consuming, laborious, and sensitive to intra- and inter-rater variability. Therefore, effective automated segmentation approaches for brain tissues are highly desirable. 

Although commonly employed segmentation methods, including SPM \cite{ashburner2005unified}, FreeSurfer \cite{fischl2002whole}, and FSL \cite{zhang2001segmentation} are effective in delineating brain tissues, they are tailored for adult MR images. These methods are less effective in segmenting infant brain MR images due to dynamic morphological and appearance changes caused by developmental processes during the first few years after birth.
To cater to the unique characteristics of infant brains, the developing human connectome project (dHCP) processing pipeline \cite{makropoulos2018developing} and Infant FreeSurfer \cite{zollei2020infant} have been developed. These are atlas-based segmentation methods, which warp manually segmented atlas to the target image using non-rigid registration methods. However, accurate segmentation of fine structures remains  difficult due to large anatomical variability, leading to sub-optimal results. Inaccurate segmentation can cause problems in the reconstruction of the cortical surfaces need to measuring cortical morphology such as cortical thickness, surface area, and curvature.

Deep neural networks (DNNs) have been successfully used for brain tissue segmentation. 
For example, Roy et al. \cite{roy2019quicknat} first pretrained a model with auxiliary labels for a large unlabeled dataset using FreeSurfer and then fine-tuned the model with limited manually labeled data. Like FreeSurfer, this method is effective only for adult images. Nie et al. \cite{nie2016fully} trained multiple models, each of which is designed for a different imaging modality to segment images particularly in isointense phase ($\approx 6-8$ months of age).
These methods are tailored for specific age groups and are not generalizable to other age groups. 

\begin{figure}[t]
\centering
\begin{subfigure}{0.2\textwidth}
  \centering
  \includegraphics[width=0.96\linewidth]{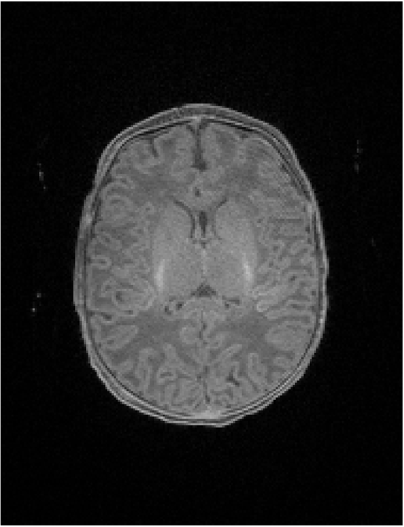}
  \caption{10 days}
  \label{fig1:sub1}
\end{subfigure}%
\begin{subfigure}{0.2\textwidth}
  \centering
  \includegraphics[width=0.96\linewidth]{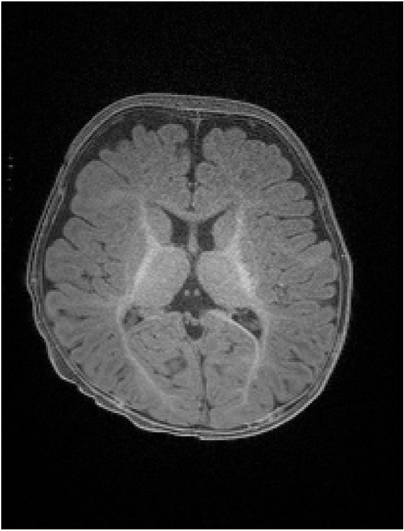}
  \caption{5 months}
  \label{fig1:sub2}
\end{subfigure}%
\begin{subfigure}{0.2\textwidth}
  \centering
  \includegraphics[width=0.96\linewidth]{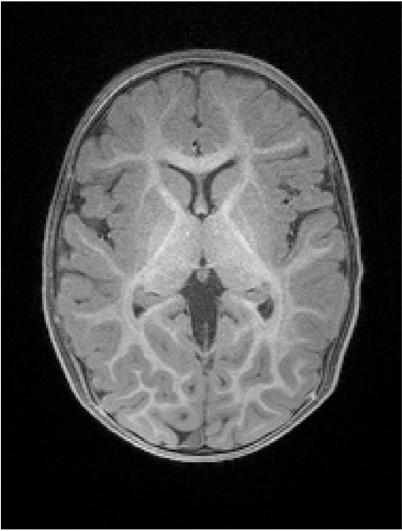}
  \caption{9 months}
  \label{fig1:sub3}
\end{subfigure}\\
\begin{subfigure}{0.2\textwidth}
  \centering
  \includegraphics[width=0.96\linewidth]{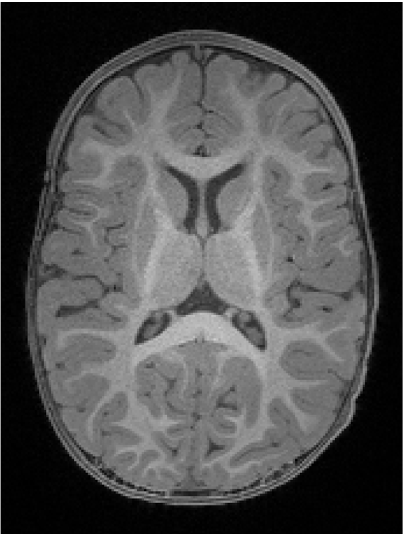}
  \caption{1 year}
  \label{fig1:sub4}
\end{subfigure}%
\begin{subfigure}{0.2\textwidth}
  \centering
  \includegraphics[width=0.96\linewidth]{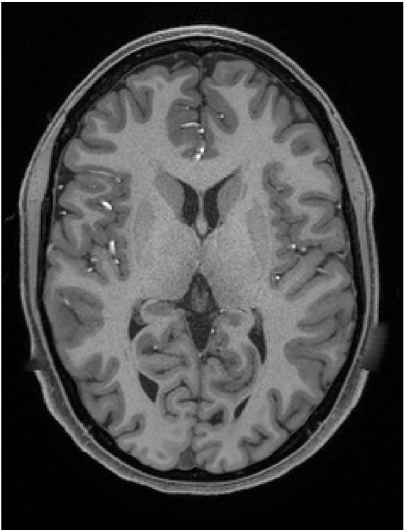}
  \caption{32 years}
  \label{fig1:sub5}
\end{subfigure}%
\begin{subfigure}{0.2\textwidth}
  \centering
  \includegraphics[width=0.96\linewidth]{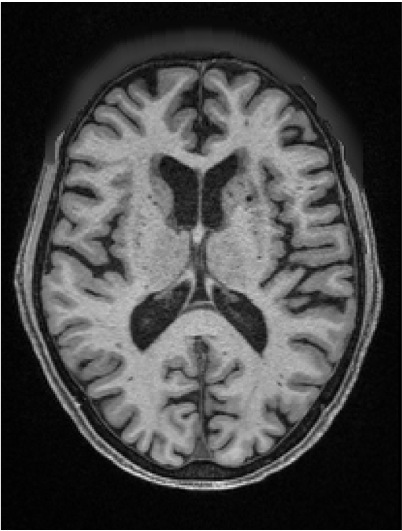}
  \caption{100 years}
  \label{fig1:sub6}
\end{subfigure}
\caption{Brain morphology and tissue contrast vary across the human lifespan.}
\label{fig1}
\end{figure}

In this paper, we propose a novel brain tissues segmentation method that is applicable across the entire human lifespan ($0-100$ years of age). Designing a unified model that takes into account the developmental and aging related structural and contrast changes in brain MR images (Fig.~\ref{fig1}) is challenging. Moreover,  common imaging issues such as noise, intensity inhomogeneity, partial volume effects, artifacts, and acquisition protocols further complicate model design. To address these problems, we propose to use supervised contrastive learning to learn discriminative feature representations that are robust to intensity differences and imaging imperfections. 

The key contributions of our work are as follows: 
\begin{inparaenum}[(i)]
  \item We propose an unprecedented lifespan segmentation method based on a single deep learning model;  
  \item We use novel supervised contrastive learning strategy to regularize network training to learn discriminative features that are robust to structural and tissue contrast changes driven by biological mechanisms, inconsistent acquisition protocols, and imaging imperfections; and 
  \item We demonstrate on a large dataset of 2,464 brain MR images the effectiveness of our method.
\end{inparaenum}

\section{Methods}

\subsection{Supervised Contrastive Learning}
Despite architecture differences, deep learning segmentation models share two main components: (i) feature extractor (FE) and (ii) classifier. For medical image segmentation, the FE usually has an encoder-decoder architecture to enable full-resolution segmentation. The FE is used to learn voxel-wise feature representations. A classifier, which is usually implemented with one or a few $1\times1\times1$ convolutional layers, is used to perform voxel-wise classification by taking the feature representations learned by the FE as the input. Previous approaches typically jointly learn the FE and the classifier via a voxel-wise loss function such as categorical cross-entropy.

However, the above joint learning scheme is unable to learn discriminative features \cite{wen2016discriminative}. As a result, the margin between feature clusters with different labels is narrow, leading to error-prone classifiers and models that are sensitive to noise, artifacts, and intensity and contrast changes. Inspired by the success of Momentum Contrast (MoCo) \cite{he2020momentum}, which is a self-supervised learning approach for instance discrimination, we adopt a similar contrastive learning strategy to learn discriminative features. Unlike MoCo, where contrastive learning is used for unsupervised pre-training, we perform contrastive learning in a supervised manner.

The key idea of contrastive learning is to pull the feature representations of the same class close together and push those from different classes far apart in a latent space. To this end, we use a memory bank to store a few typical feature representations as proxies for each semantic category. It should be noted that using proxies is more feasible than making pair-wise comparisons because around a million of voxels are involved at each training step in our case; otherwise, the problem becomes intractable. To implement contrastive learning, cosine similarity between a feature representation and the proxies in the memory bank is computed. Then, contrastive loss (i.e., InfoNCE) is calculated voxel-wise. We use the label of a given voxel as guidance to select positive and negative proxies from the memory bank and therefore we call our training strategy \emph{supervised contrastive learning}.

We use a memory bank with the shape of $C$$\times$$M$$\times$$D$, where $C$ is the number of semantic categories, $M$ is number of proxies for each category, and $D$ is the dimensionality of proxies. Note that $D$ is identical to the number of channels of the last layer of the FE. The memory bank is trainable, adaptively learned and optimized via back-propagation during training. A proxy in the memory bank acts as a positive proxy if a query (i.e., the feature representation of a voxel) shares the same label with it, otherwise it will be used as a negative proxy in calculating the contrastive loss.

The contrastive loss for an arbitrary query $\vect{q_{i}}$ is defined as:
 \begin{equation}\label{ctr_loss}
 \mathcal{L}^{i}_{\text{ctr}}=- \frac{1}{M} \sum_{\{ \vect{k^{+}_{i}} \}} \log \frac{ \texttt{sim}(\vect{q_{i}}, \vect{k^{+}_{i}}) / \tau}{\texttt{sim}(\vect{q_{i}}, \vect{k^{+}_{i}}) / \tau + \sum_{\{ \vect{ k^{-}_{i}} \}} \texttt{sim}(\vect{q_{i}}, \vect{k^{-}_{i}}) / \tau},
\end{equation}
where $\texttt{sim}(\vect{u}, \vect{v}) = \frac{\vect{u^\intercal} \vect{v}}{  \left | \vect{u} \right | \left | \vect{v} \right | }$ is cosine similarity, $\tau$ is the temperature hyper-parameter, $\vect{q_{i}}$ is the feature representation of a query voxel with index $i$, $\{\vect{k^{+}_{i}}\}$ and $\{\vect{k^{-}_{i}}\}$ are the set of positive and negative proxies for $\vect{q_{i}}$, respectively. The size of sets $\{\vect{k^{+}_{i}}\}$ and $\{\vect{k^{-}_{i}}\}$ is equal to $M$ and $M(C-1)$, respectively.

\subsection{Joint Feature Regularization and Classification}
In addition to $\mathcal{L}^{i}_{\text{ctr}}$ for feature regularization, categorical cross-entropy loss $\mathcal{L}^{i}_{\text{ce}}$ is used for joint training. The overall loss function $\mathcal{L}^{i}_{\text{total}}$ for query voxel $i$ is a weighted sum of $\mathcal{L}^{i}_{\text{ce}}$ and $\mathcal{L}^{i}_{\text{ctr}}$, balanced by $\lambda_{\text{ctr}}$, and is formally defined as
\begin{equation}
 \mathcal{L}^{i}_{\text{total}}=
\mathcal{L}^{i}_{\text{ce}} + \lambda_{\text{ctr}}\mathcal{L}^{i}_{\text{ctr}},
\end{equation}
where
\begin{equation}
 \mathcal{L}^{i}_{\text{ce}}=
-\sum_{c=1}^{C}y_{\text{ic}}\log(\hat{p}_{\text{ic}}).
\label{eq:loss}
\end{equation}
In \eqref{eq:loss}, $\hat{p}_{\text{ic}}$ is the $c$-th element of the output probability vector for query voxel $i$. $y_{\text{ic}}$, taking either 0 or 1, is the ground truth value for $\hat{p}_{\text{ic}}$.

\subsection{Learning from Imbalanced Data}
The class imbalance issue is common in brain tissue segmentation. At the image level, volumes of GM and WM are usually $2-3$ times larger than that of CSF and all the three tissues are considerably smaller than the background region, i.e., the region that is not GM, WM or CSF. At the mini-batch level, the class imbalance issue can be even more severe, since an image patch is expected to contain only voxels from as few as one class.

We solve the class imbalance issue using two different approaches. The first approach is to use stratified sampling. Instead of randomly taking samples from the whole image region, we first derive the minimum bounding box of each tissue type using the ground truth label map and then randomly take an equal number of samples (i.e., image patches) from each tissue-specific bounding box. For simplicity, we allow overlapped bounding boxes. The second approach is to select an equal number of samples from each class that is present in a mini-batch for loss computation.

Due to randomness in sampling, the label set $\mathbb{L}$ for a given mini-batch is only a subset of $\{0, 1, ..., C-1\}$ (positive integers represent foreground class labels and 0 denotes the background). We first count the number of voxels for each category in $\mathbb{L}$ and obtain the minimum $N_{\min} = \min \{ N_{c} : c \in \mathbb{L}\}$, where $N_{c}$ is the number of samples of class $c$ in the mini-batch. We then randomly select $N_{\min}$ samples from each category present in the mini-batch and compute the loss for each sample. The final loss is the average of the losses for the selected samples.

\subsection{Implementation Details}
We implemented the proposed method in Keras with TensorFlow as backend. We used a V-Net variant as the feature extractor. In contrast to the original V-Net \cite{milletari2016v}, we 
\begin{inparaenum}[(i)]
  \item removed the original head for classification; 
  \item adjusted the size of the convolutional kernel from 5 to 3 to lower the risk of overfitting; and
  \item used instance normalization \cite{ulyanov2016instance} after each convolutional layer as it is suitable for small batch sizes. 
\end{inparaenum}
We used a simple multi-layer perceptron (MLP), consisting of three $1\times1\times1$ convolutional layers, as the classifier. In the MLP, the first two convolutional layers are followed by a rectified linear unit (ReLU) and the last one is followed by a softmax layer. We used the Adam optimizer with an initial learning rate of 0.0001 to update the parameters. The convolutional kernels and the memory bank were randomly initialized. The bias terms were initialized to 0. The patch size was set to $80\times80\times80$ and the batch size to 2.

\section{Experimental Results}
\subsection{Dataset}
The dataset consists of T1- and T2-weighted image pairs from five Lifespan Human Connectome Projects \cite{Essen2012The,glasser2013minimal,zollei2020infant}: 
\begin{inparaenum}[(i)]
  \item Developing Human Connectome Project (dHCP; gestational age: 37 -- 44 weeks) \cite{makropoulos2018developing};
  \item Baby Connectome Project (BCP, ages: 0 -- 5 years) \cite{howell2019unc}; 
  \item HCP Development (HCP-D, ages: 6 -- 22 years);
  \item HCP Young Adult (HCP-YA, ages: 22 -- 37 years); and 
  \item HCP Aging (HCP-A, ages: 36 -- 100 years).
\end{inparaenum}
Of the total 2,462 image pairs used, 682 were used for training, 113 for validation and the rest for testing. Among the images used for training, we manually annotated 62 images (high-quality set) and the tissue segmentation maps for the remaining 620 images (uncorrected set) were obtained using the standard HCP minimal preprocessed pipeline. The high-quality set includes 31 images from BCP, 20 from HCP-D, 7 from HCP-YA, and 4 from HCP-D. Since the number of images in the high-quality set is much smaller than in the uncorrected set, we take half of the training samples from the high-quality set and the other half from the uncorrected set to make sure the model is not biased to uncorrected labels. For testing, 
%
%
we used dice similarity coefficient (DSC) and average surface distance (ASD) as metrics for performance evaluation.

\subsection{Experimental Setup}

We compared the proposed method with the vanilla V-Net \cite{milletari2016v} and the V-Net variant (with the same MLP as the classification head). The V-Net variant and the proposed method differ in that the V-Net variant was trained with categorical cross-entropy (CE) loss only.

\subsection{Results}
\paragraph{Main results}
The experimental results are shown in Table~\ref{tab1}. The vanilla V-Net only achieves DSC of $81.4\pm10.8$,  $85.5\pm7.6$  and $77.3\pm6.7$ on GM, WM and CSF, respectively. Compared with vanilla V-Net, the V-Net variant significantly improves the performance by $11.8\%$, $9.9\%$ and $2.8\%$ on GM, WM and CSF, respectively ($p < 0.001$). With the proposed supervised contrastive learning strategy applied, our method further improves the DSC by $1.0\%$, $0.3\%$ and $8.4\%$ on GM, WM and CSF, respectively, over the V-Net variant. Notably, the improvement on CSF is much larger than that of GM and WM because GM and WM are majority classes and can be segmented satisfactorily. However, accurate CSF segmentation is the key to ensure topological correctness in reconstructing the pial surface.

\begin{table}[t]
\centering
\caption{Comparison of DSC and ASD on the testing set for various methods.}\label{tab1}
\scriptsize
\begin{tabular}{p{0.16\textwidth}<{\centering}  p{0.1\textwidth}<{\centering} p{0.12\textwidth}<{\centering}  p{0.12\textwidth}<{\centering} p{0.12\textwidth}<{\centering} p{0.12\textwidth}<{\centering}  p{0.12\textwidth}<{\centering}}
\toprule
Method & \multicolumn{3}{c}{DSC (\%)} & \multicolumn{3}{c}{ASD (mm)} \\
\midrule
{} & \makecell[c]{GM} & \makecell[c]{WM} & \makecell[c]{CSF} & \makecell[c]{GM} & \makecell[c]{WM} & \makecell[c]{CSF} \\
V-Net \cite{milletari2016v} & $81.4\pm10.8$ & $85.5\pm7.6$ & $77.3\pm6.7$ & $0.26\pm0.11$ & $0.45\pm0.23$ & $0.28\pm0.13$\\
V-Net variant & $93.2\pm1.6$ & $95.4\pm1.2$ & $80.1\pm6.2$ & $0.08\pm0.02$ & $0.07\pm0.02$ & $0.22\pm0.12$\\
Proposed & $\textbf{94.2}\pm\textbf{1.1}$ & $\textbf{95.7}\pm\textbf{1.0}$ & $\textbf{88.5}\pm\textbf{3.8}$ & $\textbf{0.06}\pm\textbf{0.01}$ & $\textbf{0.07}\pm\textbf{0.02}$ & $\textbf{0.15}\pm\textbf{0.07}$ \\
\bottomrule
\end{tabular}
\end{table}

\paragraph{Single modality}
We studied the performance of the proposed method with only either T1- or T2-weighted image as the input. The experimental results are shown in Table~\ref{tab2}. Compared with the performance with paired T1- and T2-weighted images as input, the segmentation performance decreases marginally (ranging from $-1.9\%$ to $-0.4\%$ DSC). T1- and T2-weighted images provide complementary information that can help improve brain tissue segmentation. However, the marginal decrease indicates that our models learn effectively from limited information. 

\begin{table}[t]
\centering
\caption{Comparison of DSC and ASD on the testing set with different inputs.}\label{tab2}
\scriptsize
\begin{tabular}{p{0.16\textwidth}<{\centering}  p{0.12\textwidth}<{\centering} p{0.12\textwidth}<{\centering}  p{0.12\textwidth}<{\centering} p{0.12\textwidth}<{\centering} p{0.12\textwidth}<{\centering}  p{0.12\textwidth}<{\centering}}
\toprule
Modality & \multicolumn{3}{c}{DSC (\%)} & \multicolumn{3}{c}{ASD (mm)} \\
\midrule
{} & \makecell[c]{GM} & \makecell[c]{WM} & \makecell[c]{CSF} & \makecell[c]{GM} & \makecell[c]{WM} & \makecell[c]{CSF} \\
T1 & $92.7\pm3.0$ & $95.1\pm2.1$ & $87.2\pm6.5$ & $0.09\pm0.05$ & $0.09\pm0.06$ & $0.17\pm0.09$\\
T2 & $92.3\pm1.9$ & $94.7\pm1.8$ & $88.1\pm4.5$ & $0.09\pm0.02$ & $0.09\pm0.03$ & $0.15\pm0.06$\\
\bottomrule
\end{tabular}
\end{table}

\begin{figure}[!h] 
\centering
  \includegraphics[scale=0.22]{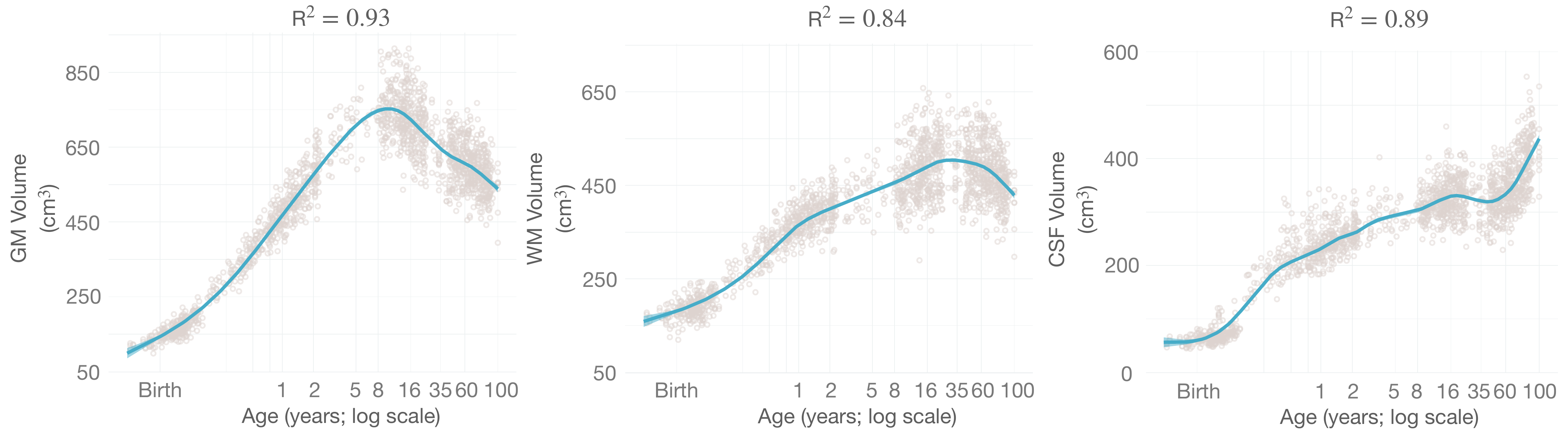}
  \caption{Growth trajectories of tissue volumes across the human lifespan.}
  \label{fig:curve}
\end{figure}

\paragraph{Growth modeling of tissue volumes}
We modeled the growth trajectories of predicted tissue volumes using generalized additive mixture model (GAMM). We fitted GAMM to the volume of each tissue type across the lifespan with cubic regression spline as smooth nonlinear function of age and subject-specific random intercept. The growth trajectories, shown in Fig.~\ref{fig:curve}, indicate that the GM and WM volumes exhibit an increase-then-decrease pattern; whereas the CSF volume increases throughout the lifespan.

\begin{figure}[ht]
	\centering
	\centerline{\includegraphics[width=0.88\textwidth]{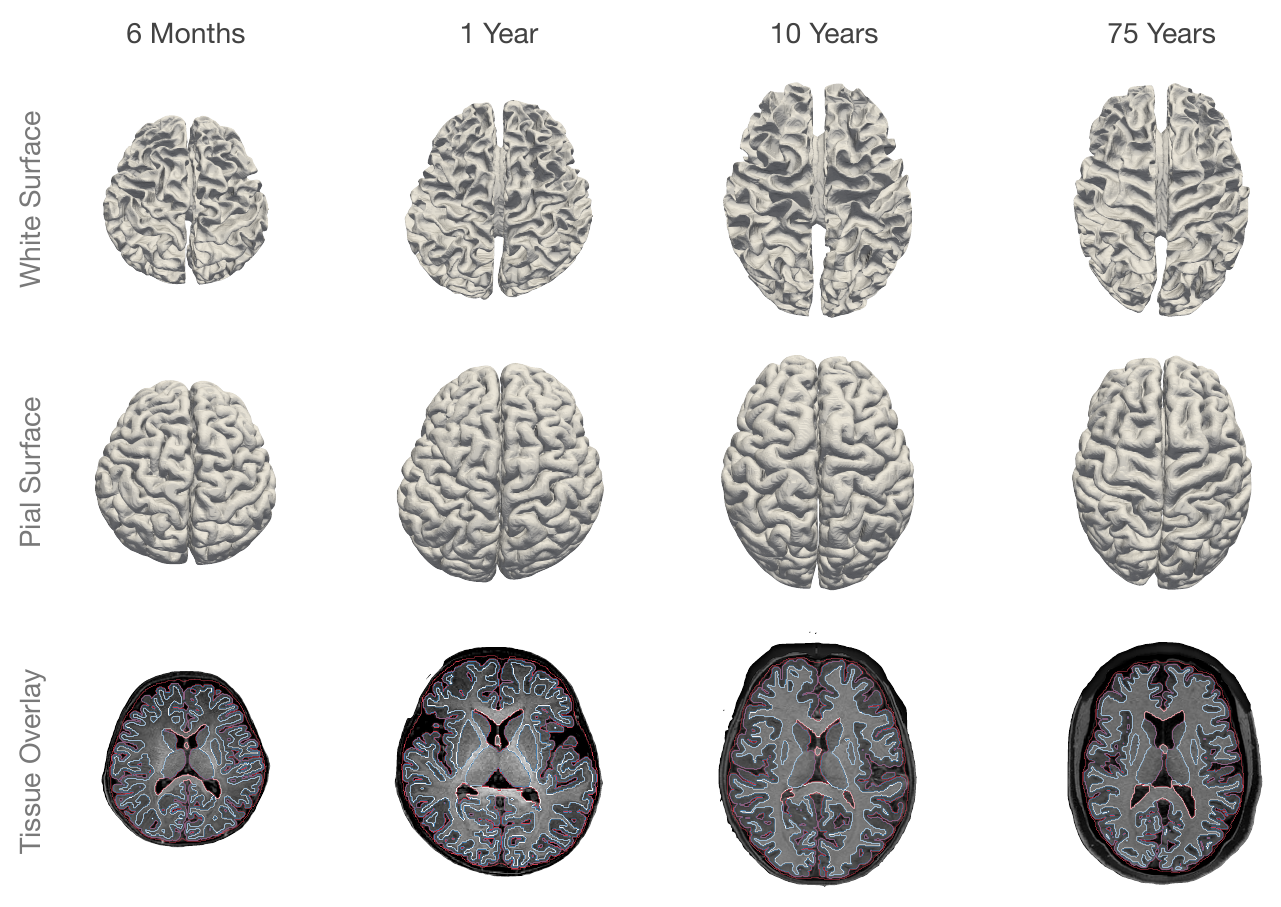}}
	\caption{Reconstructed white and pial surfaces and tissue segmentation maps at 4 different time points.} \label{Visual_results}
\end{figure}

\paragraph{Surface reconstruction} Fig.~\ref{Visual_results} shows example white/pial surfaces and tissue segmentation maps generated using our method for various time points across the human lifespan. In general, the results indicate good segmentation details particularly at the cortex as evidenced by the detailed surface convolution with clear gryi and sulci.

\subsection{Ablation Study}
\paragraph{Impact of using balanced losses} We studied the impact of selecting equal number of voxels of different categories to compute both the CE and contrastive losses. As shown in Table~\ref{tab3}, balancing the contributions from different class are critical to better segmentation of minority classes, e.g., the CSF.

\begin{table}[h]
\centering
\caption{Impact of balanced losses on DSC.}\label{tab3}
\scriptsize
\begin{tabular}{p{0.2\textwidth}<{\centering}  p{0.12\textwidth}<{\centering} p{0.12\textwidth}<{\centering}  p{0.12\textwidth}<{\centering}}
\toprule
{} & \makecell[c]{GM} & \makecell[c]{WM} & \makecell[c]{CSF} \\
Unbalanced & $93.6\pm1.5$ & $95.1\pm1.3$ & $83.0\pm6.2$\\
Balanced & $\textbf{94.2}\pm\textbf{1.1}$ & $\textbf{95.7}\pm\textbf{1.0}$ & $\textbf{88.5}\pm\textbf{3.8}$\\
\bottomrule
\end{tabular}
\end{table}

\paragraph{Impact of $\lambda_{\text{ctr}}$} We fixed $M$ to $10$ and $\tau$ to $0.1$. We tried $0.5$ and $1.0$ for $\lambda_{\text{ctr}}$. As shown in Table~\ref{tab4}, the performance is marginally changed using different $\lambda_{\text{ctr}}$.

\begin{table}[h]
\centering
\caption{Impact of $\lambda_{\text{ctr}}$ on DSC.}\label{tab4}
\scriptsize
\begin{tabular}{p{0.1\textwidth}<{\centering}  p{0.12\textwidth}<{\centering} p{0.12\textwidth}<{\centering}  p{0.12\textwidth}<{\centering} p{0.12\textwidth}<{\centering} p{0.12\textwidth}<{\centering}  p{0.12\textwidth}<{\centering}}
\toprule
{} & \multicolumn{3}{c}{0.5} & \multicolumn{3}{c}{1.0} \\
\midrule
{} & \makecell[c]{GM} & \makecell[c]{WM} & \makecell[c]{CSF} & \makecell[c]{GM} & \makecell[c]{WM} & \makecell[c]{CSF} \\
DSC & $94.0\pm5.1$ & $95.5\pm1.1$ & $\textbf{88.6}\pm\textbf{3.6}$ & $\textbf{94.2}\pm\textbf{1.1}$ & $\textbf{95.7}\pm\textbf{1.0}$ & $88.5\pm3.8$\\
\bottomrule
\end{tabular}
\end{table}

\paragraph{Impact of $M$ and $\tau$} We fixed $\lambda_{\text{ctr}}$ to 1 as it is found that the performance is insensitive to the choice of $\lambda_{\text{ctr}}$. As shown in Table~\ref{tab5}, larger $M$ and smaller $\tau$ seem generate better results. This is consistent with \cite{chen2020simple}, where it is found that a larger number of negative samples is beneficial for contrastive learning.

\begin{table}[h]
\centering
\caption{Impact of $M$ and $\tau$ on DSC.}\label{tab5}
\scriptsize
\begin{tabular}{p{0.1\textwidth}<{\centering}  p{0.12\textwidth}<{\centering} p{0.12\textwidth}<{\centering}  p{0.12\textwidth}<{\centering} p{0.12\textwidth}<{\centering} p{0.12\textwidth}<{\centering}  p{0.12\textwidth}<{\centering}}
\toprule
$\tau$ $\backslash$ $M$ &\multicolumn{3}{c}{5} & \multicolumn{3}{c}{10} \\
\midrule
0.1 & $93.9\pm5.2$ & $95.5\pm1.2$ & $88.4\pm4.0$ & $\textbf{94.2}\pm\textbf{1.1}$ & $\textbf{95.7}\pm\textbf{1.0}$ & $\textbf{88.5}\pm\textbf{3.8}$\\
0.2 & $93.6\pm5.6$ & $94.8\pm1.6$ & $88.1\pm4.2$ & $93.9\pm5.4$ & $95.3\pm1.3$ & $88.2\pm3.8$\\
\bottomrule
\end{tabular}
\end{table}

\section{Conclusion}
We have presented a novel supervised contrastive learning method for brain tissue segmentation across the human lifespan. Experimental results on a large dataset show that our method can effectively segment the tissues with paired T1- and T2-weighted images or with just T1- or T2-weighted images.

\section*{Acknowledgement}
This work was supported in part by the United States National Institutes of Health (NIH) under grants EB008374 and MH125479.
Data were provided in part by the developing Human Connectome Project, KCL-Imperial-Oxford Consortium funded by the European Research Council under the European Union Seventh Framework Programme (FP/2007-2013) / ERC Grant Agreement no. 319456. 
Data were provided in part by the Human Connectome Project, WU-Minn Consortium (Principal Investigators: David Van Essen and Kamil Ugurbil; 1U54MH091657) funded by the 16 NIH Institutes and Centers that support the NIH Blueprint for Neuroscience Research and by the McDonnell Center for Systems Neuroscience at Washington University.

%
%
%
\bibliographystyle{splncs04}
\bibliography{ref}

\end{document}